\documentstyle[pra,aps,twocolumn]{revtex}

\begin{document}
\draft
\title{\large \bf Lasing  on  the  $D_2$  line  of  sodium  in  helium
atmosphere due to optical pumping on the $D_1$ line (up-conversion)}
\author{A.A.~Apolonsky, S.A.~Babin, S.I.~Kablukov, R.V.~Markov,
\\ A.I.~Plekhanov\thanks {E-mail: fractal@iae.nsk.su},
A.M.~Shalagin.}
\address{Institute of Automation and Electrometry,
Siberian Branch of the Russian Academy of Sciences,   Novosibirsk,
630090, Russia}
\maketitle
\begin{abstract}
A  new  method  is  proposed  to  produce  population  inversion    on
transitions involving the ground  state  of  atoms.    The  method  is
realized experimentally with sodium atoms.  Lasing  at  the  frequency
corresponding to the sodium $D_2$ line is achieved in the presence  of
pump radiation resonant to the $D_1$ line with helium as a buffer gas.
\end{abstract}


In recent years,  a growing interest has been attracted to new schemes
of laser action in gases based  on  quantum  interference  induced  by
laser fields,  see review papers \cite{1,2,3,4,5,6} and  citation  therein.
Especially attractive is an opportunity to achive lasing  from  highly
excited levels into the ground state of atoms and  molecules  or  into
the  lowest  possible  energy  levels  usually  having  the    highest
population (in the absence of perturbing fields).  It is such  schemes
that are considered perspective for short-wave generation.

There  are  numerous  works investigating the ampliphication  of
radiation in alkali-metal vapors on the  transition  into  the  ground
state in the $V$-scheme. Some of  them  exploit effects of coherence to
suppress absorption on the transition from the ground-state to
appropriate upper  level while the upper level is slightly populated
by various methods such  as  discharge   current
\cite{Gao_1992}, collisions with buffer gas particles \cite{Kleinfeld}.

In this paper we aim at drawing  attention  to  new  possibilities  to
produce laser action on transitions into the ground state.  The method
is based on manipulation with polarization of pump waves and  specific
collisional population transfer mechanism.

In earlier  papers  \cite{Glushko,Atutov}  the  possibility  of  laser
action was demonstrated at the frequency corresponding  to  the  $D_1$
line in the presence of strong pump field tuned in resonance with  the
$D_2$ line in  alkali-metal  vapors.    The  effect  is  generated  by
collisions with  buffer-gas  particles.    The  collisions  should  be
frequent enough  in  order  to  establish  Boltzmann  distribution  of
population between fine-structure components ($P_{3/2}$ and $P_{1/2}$)
during pump pulse action.    At these  conditions  magnetic  sublevels
populations of the $P_{1/2}$ state appear  by  the  Boltzmann  factor
higher than that of the $P_{3/2}$ state.  A strong pump field is  used
to equalize  magnetic sublevels  populations  of  the  ground  state
$S_{1/2}$ and the resonant upper state $P_{3/2}$.  As a result,    the
population of level $P_{1/2}$ appears to be higher than the population
of level $S_{1/2}$. Thus,  the population inversion on transition from
the upper state $P_{1/2}$ into the ground state $S_{1/2}$ is achieved,
and laser action at the frequency of the $D_1$ line remains  possible.
In \cite{Glushko,Atutov} lasing reached the superluminosity regime.

In correspondence with nature  of the process the generated  frequency
in \cite{Glushko,Atutov} was lower  compared  to  the  pump  frequency
(down-conversion).  This naturally  leads  to  the  question:  whether
up-conversion is possible, i.e.  to achieve laser action (on the $D_2$
line) with frequency higher then that for pump field (resonant to  the
$D_1$ line)?  The present paper gives a  positive  answer.    Such  an
opportunity appears if one combines the  above  mentioned  collisional
processes with  specific  polarization  effects.    Let's  prove  this
statement.  Consider an interaction of the  pump  pulse  (specifically
polarized) with a gas of atoms in mixture with a buffer gas at a  high
pressure.  Let's take for definitness sodium atoms with  corresponding
level scheme, see Fig.\ref{f1}.
The pump pulse with carrier frequency resonant to the $D_1$  line  has
the following temporal and polarization structure. It consists of long
low-intensity circularly-polarized prepulse and  much  more  intensive
and short main pulse  with  orthogonal  circular  polarization.    The
prepulse is used for optical orientation of sodium atoms in the ground
state.  From this one can set the requirement on its duration (it  has
to be longer compared to upper state relaxation  time)  and  intensity
limit (it may be low, but enough for optical orientation only).  After
the end of the prepulse almost all the population is optically  pumped
into one of the magnetic sublevels of the ground state (its population
is shown in Fig.\ref{f1}(a) by symbolic column).  The main pulse  may
be shorter than the  relaxation  time  of  the  excited  levels.    It
transfers the population from the ground state onto  sublevel  of  the
excited  level  $P_{1/2}$  that  is  initially  empty   (corresponding
transition is shown in  Fig.\ref{f1}(b)  by  a  solid  arrow).    The
intensity of the main pulse must be high enough to  maintain  equality
of populations for the coupled  sublevels.    Furthermore,    at  high
buffer-gas pressure collisions are frequent enough to mix  the  states
$P_{1/2}$ and $P_{3/2}$ and their magnetic sublevels as well.    As  a
result the population is  distributed  between  the  sublevels  almost
equally,  as shown in Fig.\ref{f1}(b) by small columns.  The sublevel
$M=-1/2$ of the ground state is almost  empty  (remember  that  it  is
pumped out by the prepulse),  whereas other sublevels  of  the  ground
state and excited states are populated almost equally (the  population
difference by the Boltzmann factor  between  $P_{1/2}$  and  $P_{3/2}$
states is insignificant here and may be neglected).  One can see  that
population inversion between some upper magnetic sublevels  (including
those for the $P_{3/2}$ level) and the sublevel $M=-1/2$ of the ground
state is created.  Hence,   necessary  conditions  for  laser  action,
specifically from the level $P_{3/2}$ to the  ground  state  $S_{1/2}$
are met.

Thus,  we have shown that there exist conditions for  producing  laser
action on the $D_2$ line with pulsed excitation resonant to the  $D_1$
line.  Taking into account partial oscillator strengths,  we see  that
the highest  gain  is  achieved  on  the  transition  $P_{3/2}(M=-3/2)
\rightarrow S_{1/2}(M=-1/2)$,  as shown in Fig.\ref{f1}(b) by a  wavy
arrow.  It  means  that the generated wave  has  presumably  the  same
polarization as the main pump pulse.  To avoid misunderstanding,    we
should remind that collisions with particles  of  non-magnetic  buffer
gas (noble gas as an example) mix magnetic  sublevels  of  the  ground
state in alkali metals very weakly,  so we neglect  this  factor  with
confidence.

On the basis of the above treatment,  we have experimentally  realized
such  scheme with population  inversion  on   transition    $3P_{3/2}-
3S_{1/2}$ of sodium.  A general schematic of our experimental setup is
shown in Fig.\ref{f2}.
In the experiment  it  was  easier  and  more  convenient  to  use  CW
radiation instead of the prepulse.  In  this  case  it  maintains  the
orientation of the ground state between  pump  pulses,    coming  from
another laser source.  We used CW dye laser $DL1$ with linear vertical
polarization (of $\approx3$~GHz spectral  linewidth)  and  pulsed  dye
laser $DL2$ with horizontal  polarization  (of  10~GHz  linewidth  and
$\approx5$~ns pulse duration).  In the spectrum of  the  pulsed  laser
together with narrow-band laser radiation, a broadband luminescence of
the R6G dye is present,  but its spectral density is by  3  orders  of
magnitude weaker.

After passing the quarter-wave ($\lambda/4$) plate the beam of the  CW
laser becomes clockwise polarized,  and the beam of the
pulsed laser is counter-clockwise polarized.    The  CW
radiation prepares the medium converting it into the  $S_{1/2}(M=1/2)$
state,  and high-power pulses are used for  populating  upper  levels.
Both beams propagate in the same direction and are focused by the lens
$L_1$ with focal length $F=55$~cm into  the  cell  with  sodium-helium
mixture. The intensity near the beam waist is $\approx60$~W/cm$^2$ for
the CW laser,  and up to $6$~MW/cm$^2$ for the pulsed laser.  The cell
has 1.5-cm diameter and is 22-cm  long  with  heated  zone  ($BC$)  of
4.5~cm in the central part. The sodium vapour density is controlled by
varying the temperature, measured by thermocouple.  The cell is placed
between the Helmholz coils ($HC$)  that  provide  external  longitudinal
magnetic  field  to  eliminate  deorientating  effect  of   transverse
component of the laboratory field. The magnitude of the external field
up to $B \approx80$~G is available. Output radiation is focused by the
lens $L_2$ onto the slit of the monochromator  RAMANOR~HG.2S~($M$)  with
an apparatus width of about 0.5~cm$^{-1}$.  Data from  photomultiplier
$D$ connected to  amplifier  and  integrator  are  registered  with  a
computer.  That allows us to store and average  measured  data.    The
generation at the $D_2$ line frequency was measured  in  direction  of
the pump beam as well as in the opposite direction.  For this  purpose
the beam splitter ($BS_2$) was inserted into the pump beam pathway, as
shown in Fig.\ref{f2}.

First of all,  it has been ascertained,  that in the  absence  of  the
external magnetic field and at low buffer-gas (helium) pressure  there
is no coherent radiation at the $D_2$ line frequency in a broad  range
of other experimental parameters.  After the external  magnetic  field
$B>0.5$~G is applied,  an intense coherent radiation at the $D_2$ line
frequency appeares at helium pressure higher than 200~torr with CW and
pulse lasers being tuned to exact resonance with the $D_1$ line,   see
curve~1 in Fig.\ref{f3}(a).  Divergence of the output beam  appeares
to be no more than that of the pump beam. Registered spectral width is
about $0.7$~cm$^{-1}$,  being close  to  the  apparatus  width.    The
radiation at the $D_2$ line frequency has nearly the same polarization
as the strong pulsed field.  Curve~2 in Fig.\ref{f3}(a)  illustrates
the measurement without magnetic field ($B=0$).  In this case only the
absorption line is observed because broadband R6G dye luminescence  is
absorbed in optically thick media.
The coherent radiation at the $D_2$ line frequency is observed both in
the direction of the  pump  beam,    and  in  the  opposite  direction
(Fig.\ref{f3}(b)).  Spectral width for the forward and backward output
radiations is nearly the same.  In contrast to the forward  radiation,
the backward one is also observable in the  absence  of  the  external
magnetic field.  Note,  that  the  intensity  of  the  forward  output
radiaton is 80 times as high as that of the backward radiation (in the
presence of magnetic field).  This fact can be explained by the effect
of dye luminescence that serves as a seed leading to amplification  in
its direction.

It is interesting to note that the backward generation occurs  in  the
absence of longitudinal magnetic  field,    i.e.    when  the  optical
orientation is noticeably destroyed.  We suppose that this happens due
to a sufficient intensity of the CW radiation that is able to transfer
the residual population from sublevel $S_{1/2}$ $M=-1/2$ into  excited
states,  thus helping to create inversion on the operating transition.
Absence of forward  generation  under  identical  conditions  is  most
likely  explained  as  follows.    The  CW  radiation  gets   absorbed
propagating along the heated  zone.    Because  of  this,    inversion
condition is no longer valid in the output part of the zone.    Hence,
the generated radiation coming forward is absorbed in the output part.

The  output  intensity  of  generated  radiation  under  fixed   other
conditions reaches its maximum when frequencies of both lasers (CW and
pulsed) $\omega_L$ are tuned in  resonance  with  the  $D_1$  line  of
frequency $\omega_{D_1}$. Detuning from the exact resonance $|\Omega| =
|\omega_L-\omega_{D_1}| \approx4$~cm$^{-1}$ leads to disappearance  of
the output signal.  Appearing at helium pressure  of  200~torr,    the
intensity  at  the  $D_2$  line  frequency  rises  monotonically  with
increasing pressure  up  to  810  torr,    highest  available  in  the
experiment.  The maximum of output intensity measured as a function of
sodium  vapour  density    is    reached    at    $N\approx8    \times
10^{12}$~cm$^{-3}$.  Under these conditions the cell  transmission  is
$\approx90$\% for the pulse radiation,  and $\approx80$\% for  the  CW
laser beam.

Starting at external magnetic field  strength  $B$  as  low  as  0.5~G
(comparable to laboratory field),   the  output  signal  grows  almost
linearly with increasing field up to $B\approx5$~G when it  saturates.
Note,  that an application of the external magnetic field considerably
helps to orientate sodium atoms by circularly polarized CW  radiation.
The luminescence intensity is attenuated by a factor of $3\div4$ after
the field is applied, that means the portion of oriented atoms reaches
more than 80\%.

When the pump intensity is attenuated down to $1.5  \div2$~MW/cm$^{2}$
the  output  intensity  does  not  change  significantly.      Further
attenuation of the pump radiation leads to a smooth  decrease  in  the
generated power while its spectral width remaines constant.    At  the
same time,  attenuation of the CW laser power leads to an abrupt  fall
in the generated power at  the  $D_2$  line  frequency.    At  optimal
conditions ($\Omega=0$,  $N\approx10^{13}$~cm$^{-3}$,  helium pressure
$\approx810$~torr,  $B\approx5$~G) the output intensity  of  generated
radiaton reaches $1.5\div2$\% of the absorbed pump intensity  (of  the
pulsed laser).  The intensity of the generated  radiation  inside  the
cell is estimated to be $\sim 5$~kW/cm$^{2}$, corresponding to  strong
saturation,  $\ae \gg1$\footnote{Saturation parameter  is  defined  as
$\ae = \left( {d_{mn} E/ 2 \hbar} \right)^2/ \Gamma \Gamma_m$,   where
$d_{mn}$ -- matrix element of the dipole moment on  transition  $m-n$,
$E$ -- electric field amplitude,  $\Gamma$ --  collisional  linewidth,
$\Gamma_m$ -- relaxation rate (radiative) of the  upper  level  $m$.}.
This suggests that the generation occurs in superluminosity regime (at
least in the forward direction), i.e. the generated wave is capable to
utilize considerable portion of inversion on the operating transition.

We also have considered parametric processes such as wavemixing as
alternative interpretation of our experimental results.
It is well known that wavemixing processes are suppressed at the resonance
conditions due to absorption \cite{Harter}, while our signal 
conversely has maximum at the frequensy resonant to the $D_2$ line 
and vastly decays with detuning from the resonance, and moreover,
signal generated in wavemixing in backward direction not be present
due to phase mismatching \cite{Kirin,Arutyunian}. 
Thus our original interpretation presented above seems to be
more realistic and reasonable.

Our work is based on the composition of well known 
phenomena (laser excitation, optical pumping, collisional 
population transfer), but we need to stress that only 
appropriate combination of the phenomena mentioned above
lead to the new results which were obvious beforehand. 

Thus, we have shown in the present work,  that proposed combination of
polarization  and  collisional  transfer    mechanisms    opens    new
opportunities for resonant  radiative  processes.    In  view  of  the
development of the general method applied here for special case,    we
propose a variant of the scheme that allows one to achieve  generation
of violet and UV radiation on transitions into the ground state.   For
this purpose,  instead  of  a  single-photon  process  a  two-step  or
two-photon excitation into higher-lying states can be  explored
(see Fig.\ref{f4}).    In
this case,  radiative transition from the excited state ($m$) into the
ground state is parity-forbidden.  However,  if there is another close
(within  $k  T$  range)  level  $l$  of  different  parity,    coupled
radiatively with the ground state $n$,  the developed  method  remains
applicable.  Using high-pressure buffer gas an  efficient  collisional
mixing of upper levels $m$ and $l$ can be provided.  Then,  due to the
Boltzmann factor (if level $l$ is  lower  than  level  $m$)  and  high
intensity pump waves,  a population inversion on transition $l-n$  can
be achieved,  thus resulting in laser action in short-wave range.   In
case the available intensity and the Boltzmann factor are  not  enough
for generation,  or level $l$ lies higher  than  level  $m$,    it  is
possible to prepare the system using the same polarization  technique:
optical orientation of the ground state $n$ by resonant radiation  (at
first-step transition) opens an additional  opportunity  to  build  up
population inversion between  corresponding  sublevels  of  transition
$l-n$.

We are gratefully acknowledge fruitful discussions with  S.G.~Rautian,
Ye.V.~Podivilov and M.G.~Stepanov,  and an opportunity to use  CW  dye
laser by "INVERSion" and "Technoskan"  companies.    This  work  was
supported in part by Russian Foundation for Basic  Research,    grants
96-15-96642, 98-02-17924.

\begin{figure}
\caption{Relevant  energy  levels  of  Na  and  optical   transitions
(quantization axis is collinear to wave vectors).}
\label{f1}
\end{figure}

\begin{figure}
\caption{Schematic of the experiment to observe superluminosity on the
sodium $D_2$ line. $DL$, dye lasers;  $P$,  polarizers;  $L$,  lenses;
$HC$,  Helmholtz coils;  $BC$,  bifilar heater  coil;    $BS$,    beam
splitters; $M$, monochromator; $D$, detector.}
\label{f2}
\end{figure}

\begin{figure}
\caption{
Output lasing signal around the $D_2$ line frequency  $(\omega_{D_2})$
viewed from opposite directions.  Forward signal on the $D_2$ line (a)
is 80 times higher then the backward signal (b).  Curves 1 (a,b)  show
output signal of the $D_2$ line with $D_1$ pumping for 810 torr helium
pressure,  $T=495$ K,  $B=80$ G.  Curves 2 (a,b) show output signal of
the $D_2$ line at the same condition but for $B=0$. Pump beams detunings
from the $D_1$ resonance are equal to zero.}
\label{f3}
\end{figure}

\begin{figure}
\caption{
Elucidation of the posibility of shortwave lasing.}
\label{f4}
\end{figure}

\end{document}